\documentstyle[11pt,twoside,newpasp,epsf,natbib]{article}
\def\edcomment#1{\iffalse\marginpar{\raggedright\sl#1\/}\else\relax\fi}
\reversemarginpar

\newcommand{\mem}[1]{\mathrm{ #1}}
\newcommand{\n}{\mem{n}}
\newcommand{\p}{\mem{p}}

\newcommand{\cdr}{^{13}\mem{C}}

\newcommand{\nvi}{^{14}\mem{N}}

\newcommand{\cvi}{^{14}\mem{C}}
\newcommand{\nfu}{^{15}\mem{N}}

\newcommand{\jahre}{\, \mathrm{yr}}

\begin{document}
\title{Could SiC A+B grains have originated in a post-AGB thermal
pulse?}
 \author{Falk Herwig}
\affil{University of Victoria, Victoria, BC, Canada}
\author{Sachiko Amari}
\affil{Washington University, St.\ Louis, USA}
\author{Maria Lugaro}
\affil{IoA, University of Cambridge, Cambridge, UK}
\author{Ernst Zinner}
\affil{Washington University, St.\ Louis, USA}

\begin{abstract}
The carbon and nitrogen isotopic ratios of pre-solar SiC grains of type A+B
suggest a proton-limited nucleosynthetic process as encountered, for instance,
during the very late thermal pulse of post-AGB stars. We study the nuclear
processes during this phase and find carbon and nitrogen isotopic ratios which
can reproduce those of A+B grains. These results are still preliminary
because they depend on uncertain factors such as the details of
mixing during the post-AGB
thermal pulse, the rates of some nuclear reactions, and the
assumptions on mixing during the progenitor AGB phase.
\end{abstract}

In recent years isotopic analysis of individual pre-solar SiC grains found
in primitive meteorites have established various sub-groups and linked these
to different stellar production sites (Zinner, 1998).
However, the origin of SiC A+B grains remains elusive. These grains are mainly
characterized by low $^{12}$C/$^{13}$C ratios (2-10) and a wide range of values
for the $^{14}$N/$^{15}$N ratio (10$^2$-10$^4$) (Fig.\,1, left panel), while
showing close to solar silicon isotopic ratios (Amari et al., 2001). Some
stellar
sources, such as novae or J-type Asymptotic Giant Branch (AGB) stars,
have been considered but cannot explain all the isotopic ratios. These
ratios suggest a proton-limited nucleosynthetic process. The conditions for
such a process are encountered when small amounts of protons are ingested
into a hot carbon-rich region. One stellar evolution phase during
which these conditions are met is the very late thermal pulse (VLTP) that
occurs in roughly 10$\%$ of all central stars of Planetary Nebulae
(Herwig et al., 1999; Herwig, 2001).

We use a fully implicit code to produce one-zone nucleosynthesis
models for conditions typical of the VLTP event. The models discussed here
are the
\begin{figure}
\plottwo{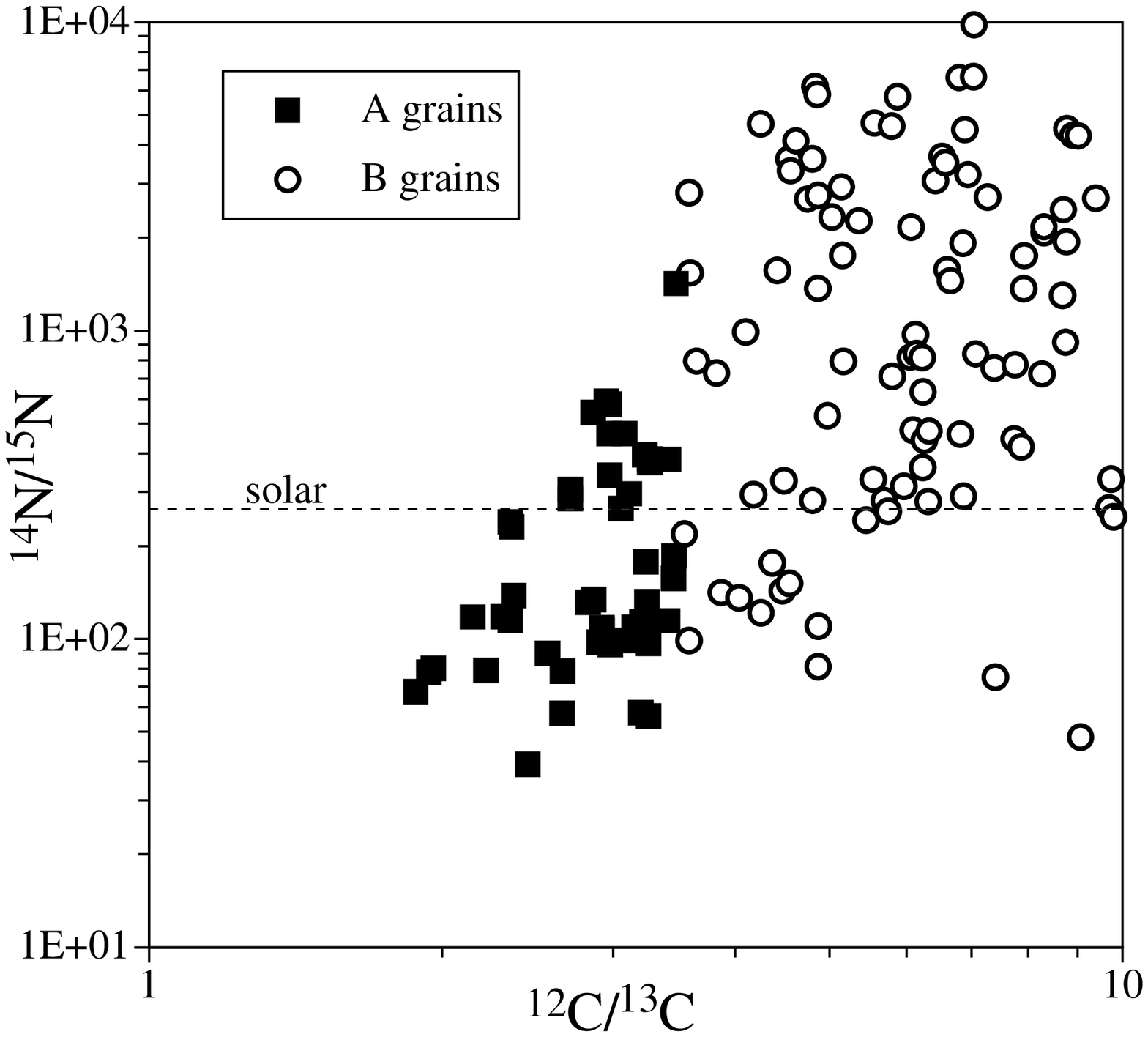}{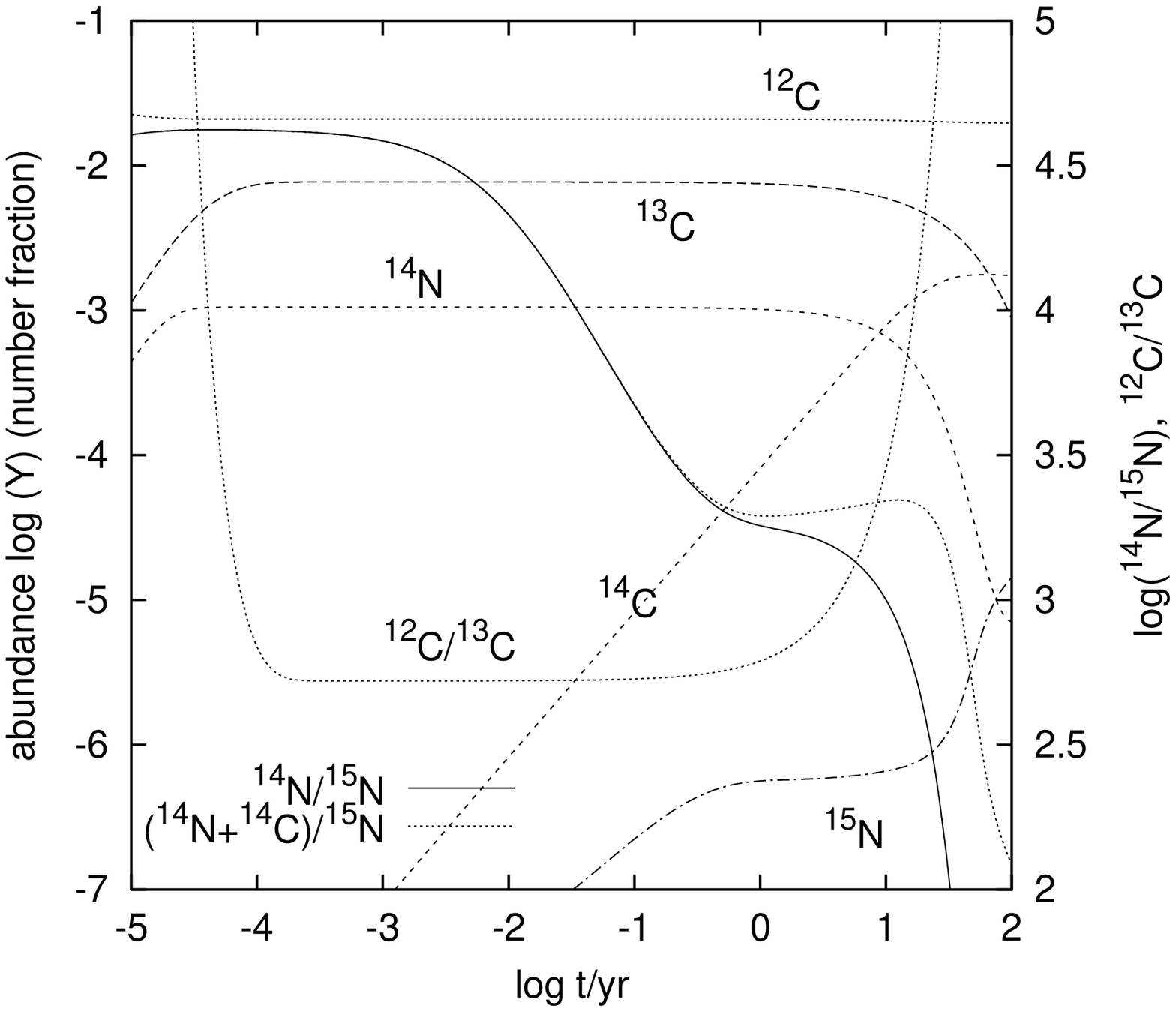}
\caption{\textbf{Left:} Carbon and nitrogen isotopic ratios of A+B SiC
grains. The C isotopic ratios of some grains are smaller than expected from
CNO-cycle
equilibrium. \textbf{Right:} Evolution of isotopic abundances according to a
one-zone model for typical conditions encountered during the VLTP H-ingestion
phase  ($T_8=1.25$, $X(H)_{\rm initial}$=0.01).}
\end{figure}
same as those presented in Herwig, Lugaro \& Werner (this conference). The
abundance variations are shown in Fig.\,1 (right panel).
Due to the large $^{12}$C/H ratio ($\sim 50$) most protons lead to the
production
of $^{13}$C while few are available for the synthesis of a smaller amount of
$^{14}$N. Neutrons released by the $\cdr(\alpha,\n)$ reaction
 are captured predominantly by $\nvi$. Neutron captures by this isotope
create $\cvi$ via $(\n,\p)$ and  $\nfu$ through $(\n,\gamma)$. After $t =
1\jahre$ $\nfu$ is
destroyed by the $(\p,\alpha)$ reaction. The $\nvi$/$\nfu$ ratio is very
sensitive
to this and to the $\nvi(\n,\p)\cvi$ reaction.
The final results depend on the chosen temperature as well as on the
amount of hydrogen assumed to enter the burning layer.
These quantities
depend on the mass and metallicity of the parent star and on the  details
of convective mixing which are still not well known.
 For example, the $T_8=1.35$,
X(H)=0.008 model results after a time of 0.1yr in
$^{12}$C/$^{13}$C$\sim$2.5 and $^{14}$N/$^{15}$N$\sim$1000.
Another source of uncertainty is introduced by the choice of the initial
abundances, which depend on the assumptions about mixing during the
progenitor AGB phase.
Our models can be extended to
evolutionary phases of other objects such as accreting white dwarfs, and
we plan to include other isotopic ratios in the study, in particular those of
silicon. \\

\noindent
\textbf{Acknowledgements:} F.H. would like to thank D.A. VandenBerg for support
through his Operating Grant from the Natural Science and Engineering Research
Council of Canada.

\end{document}